# The role of decomposition reactions in assessing first-principles predictions of solid stability

Christopher J. Bartel[1], Alan W. Weimer[1], Stephan Lany[2], Charles B. Musgrave[1,2,3*], Aaron M. Holder[1,2*]

[1]Department of Chemical and Biological Engineering, University of Colorado, Boulder, Colorado 80309, USA
[2]National Renewable Energy Laboratory, Golden, CO 80401, USA
[3]Department of Chemistry, University of Colorado, Boulder, Colorado 80309, USA

[*]Correspondence to: charles.musgrave@colorado.edu, aaron.holder@colorado.edu

**ABSTRACT**

The performance of density functional theory (DFT) approximations for predicting materials thermodynamics is typically assessed by comparing calculated and experimentally determined enthalpies of formation from elemental phases, $\Delta H_f$. However, a compound competes thermodynamically with both other compounds and their constituent elemental forms, and thus, the enthalpies of the decomposition reactions to these competing phases, $\Delta H_d$, determine thermodynamic stability. We evaluated the phase diagrams for 56,791 compounds to classify decomposition reactions into three types: 1. those that produce elemental phases, 2. those that produce compounds, and 3. those that produce both. This analysis shows that the decomposition into elemental forms is rarely the competing reaction that determines compound stability and that approximately two-thirds of decomposition reactions involve no elemental phases. Using experimentally reported formation enthalpies for 1,012 solid compounds, we assess the accuracy of the generalized gradient approximation (GGA) (PBE) and meta-GGA (SCAN) density functionals for predicting compound stability. For 646 decomposition reactions that are not trivially the formation reaction, PBE [mean absolute difference between theory and experiment (MAD) = 70 meV/atom] and SCAN (MAD = 59 meV/atom) perform similarly, and commonly employed correction schemes using fitted elemental reference energies make only a negligible improvement (~2 meV/atom). Furthermore, for 231 reactions involving only compounds (Type 2), the agreement between SCAN, PBE, and experiment is within ~35 meV/atom and is thus comparable to the magnitude of experimental uncertainty.

# INTRODUCTION

The design and discovery of new materials are being rapidly accelerated by the growing availability of density functional theory (DFT) calculated property data in open materials databases, which allow users to systematically retrieve computed results for experimentally known and yet-to-be-realized solid compounds.[1-6] The primary properties of interest are the optimized structure and corresponding total energy, $E$, with, for example, ~50,000,000 compiled structures and energies available *via* the NOMAD repository.[7] Given $E$ for a set of structures, one can routinely obtain the reaction energy, $E_{rxn}$, to convert between structures. $E$ for a compound is typically compared with $E$ for its constituent elements to obtain the formation enthalpy, $\Delta H_f$, which provides the thermodynamic driving force at zero temperature and pressure for stability of a given structure with respect to its constituent elements:

$$\Delta H_{f, A_{\alpha_1} B_{\alpha_2} \ldots} = E_{A_{\alpha_1} B_{\alpha_2} \ldots} - \sum_i \alpha_i E_i \qquad [1]$$

where $E$ is the calculated total energy of the compound ($A_{\alpha 1}B_{\alpha 2}\ldots$), $\alpha_i$ the stoichiometric coefficient of element $i$ in the compound, and $E_i$ the total energy (chemical potential) of element $i$. $\Delta H_f$ computed by **Equation 1** is typically compared to $\Delta H_f$ obtained experimentally at 298 K with varying degrees of agreement depending on the density functional and compounds (chemistries) under investigation.[2,3,8-13]

However, $\Delta H_f$ is rarely the useful quantity for evaluating the stability of a compound. The reaction energy for a given compound relative to all other compounds within the same composition space is a more relevant metric for accessing stability, where the reaction with the most positive $E_{rxn}$ is the decomposition reaction.[11,14,15] For example, for a given ternary compound, $ABC$, the relevant space of competing materials includes the elements ($A$, $B$, and $C$), all binary compounds in the $A$-$B$, $A$-$C$, and $B$-$C$ spaces, and all ternary compounds in the $A$-$B$-$C$ space. The stability of $ABC$ is obtained by comparing the energy of $ABC$ with that of the linear combination of competing compounds with the same average composition as $ABC$ that minimizes the combined energy of the competing compounds, $E_{A-B-C}$. The decomposition enthalpy, $\Delta H_d$, is then obtained by:

$$\Delta H_d = E_{rxn} = E_{ABC} - E_{A-B-C}. \qquad [2]$$

$\Delta H_d > 0$ indicates an endothermic reaction for a given compound $ABC$ forming from the space of competing compounds, $A$-$B$-$C$; the sign notation that $\Delta H_d > 0$ indicates instability is chosen to be commensurate with the commonly reported quantity, "energy above the hull", where $\Delta H_d$ also provides the energy with respect to the convex hull but can be positive (for unstable compounds) or negative (for stable compounds). A ternary example was shown for simplicity, but the decomposition reaction and $\Delta H_d$ can be obtained for any arbitrary compound comprised of $N$ elements by solving the $N$-dimensional convex hull problem.

For the high-throughput screening of new materials for a target application, stability against all competing compounds is an essential requirement for determining the viability of a candidate material.[15] In this approach, compounds are typically retained for further evaluation (more rigorous calculations or experiments) if $\Delta H_d < \gamma$, where the threshold $\gamma$ commonly ranges from ~20 to ~200 meV/atom depending on the priorities of the screening approach and the breadth of materials under evaluation.[16-21] The success of high-throughput screening approaches thus depends directly on the accuracy of $\Delta H_d$, which is typically obtained using DFT with routinely employed approximations to the exchange-correlation energy. Nevertheless, despite the intimate link between stability predictions and $\Delta H_d$, new approaches (e.g., the development of improved density functionals and/or statistical correction schemes) are primarily benchmarked against experimentally obtained $\Delta H_f$. Here, we show that the decomposition reactions that are relevant to stability can be classified into three types, and that the ability of DFT-based approaches to predict $\Delta H_d$ for each type relative to experiment is the appropriate determinant of the viability of that method for high-throughput predictions of compound stability.

## RESULTS
**Relevant reactions for determining the stability of compounds**

The decomposition reactions that determine $\Delta H_d$ fall into one of three types: Type 1 – a given compound is the only known compound in that composition space, the decomposition products are the elements, and thus $\Delta H_d = \Delta H_f$ (**Fig. 1**, left); Type 2 – a given compound is bracketed (on the phase diagram) by compounds and the decomposition products are exclusively these compounds (**Fig. 1**, center); and Type 3 – a given compound is not the only known compound in the composition space, is not bracketed by compounds and the decomposition products are a combination of compounds and elements (**Fig. 1**, right). For a given compound, one of these three types of decomposition reactions will be the relevant reaction for evaluating that material's stability. Notably, these decomposition reactions apply to both compounds that are stable (vertices on the convex hull, $\Delta H_d \leq 0$, **Fig. 1**, top) and unstable (above the convex hull, $\Delta H_d > 0$, **Fig. 1**, bottom).

As it pertains to thermodynamic control of synthesis, Type 2 reactions are insensitive to adjustments in elemental chemical potentials that are sometimes modulated by sputtering, partial pressure adjustments, or plasma cracking. Any changes to the elemental energies will affect the decomposition products and the compound of interest proportionally, and therefore, while $\Delta H_f$ for all compounds will change, $\Delta H_d$ will be fixed. This is in contrast to Type 1 reactions which become more favorable with increases in the chemical potential of either element. The thermodynamics of Type 3 reactions can be modulated by these synthesis approaches if the elemental form of the species whose chemical potential is being adjusted participates in the decomposition reaction, i.e. the compound must be the nearest (within the convex hull construction) stable, or lowest energy metastable, compound to the element whose chemical potential is being adjusted.[22]

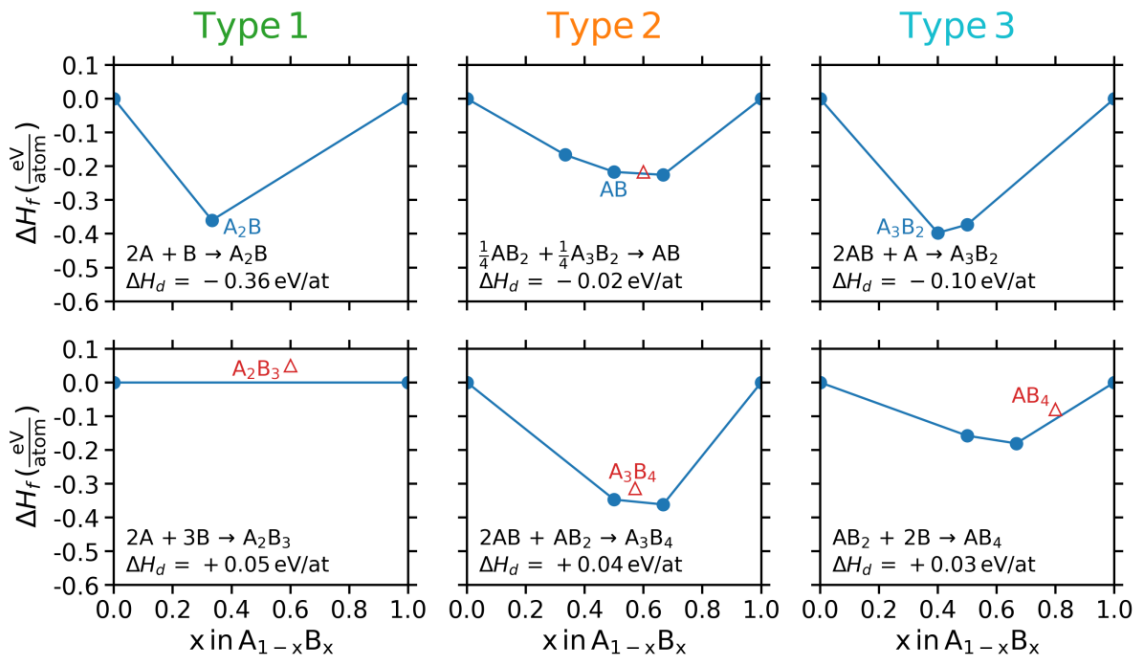

**Figure 1. Three unique decomposition reactions** A stable (top) and metastable (bottom) example of each reaction type. Left: reaction Type 1 – the decomposition products are the elements; Center: reaction Type 2 – the decomposition products contain no elements; Right: reaction Type 3 – the decomposition products contain elements and compounds. Solid blue circles are breaks in the hull (stable) and open red triangles are above the hull (metastable). In all examples, A and B are arbitrary elements. We note that in the stable Type 2 example (top center), the stability of AB is determined by a stable compound, $AB_2$ and an unstable compound, $A_3B_2$. This particular phase diagram is chosen to emphasize that the decomposition of stable compounds can include unstable compounds.

The relative prevalence of each decomposition pathway is not yet known, although the phase diagrams of most inorganic crystals can be resolved using open materials databases. At present, the Materials Project[1] provides 56,791 unique inorganic crystalline solid compounds with computed $\Delta H_f$. Using the $N$-dimensional convex hull construction, we determined $\Delta H_d$ and the stability-relevant decomposition reaction for each compound and report the prevalence of each reaction type in **Fig. 2**. For these 56,791 compounds, Type 2 decompositions are found to be most prevalent (63% of compounds) followed by Type 3 (34%) and Type 1 (3%) decompositions. Notably, 81% of Type 1 reactions (for which $\Delta H_d = \Delta H_f$) are for binary compounds, which comprise only 13% of compounds tabulated in Materials Project. In contrast, < 1% of the non-binary compounds compete for stability exclusively with elements (**Fig. 2**, right). As the number of unique elements in the compound, $N$, increases it becomes increasingly probable that other compounds will be present on the phase diagram and the decomposition will therefore be dictated by these compounds.

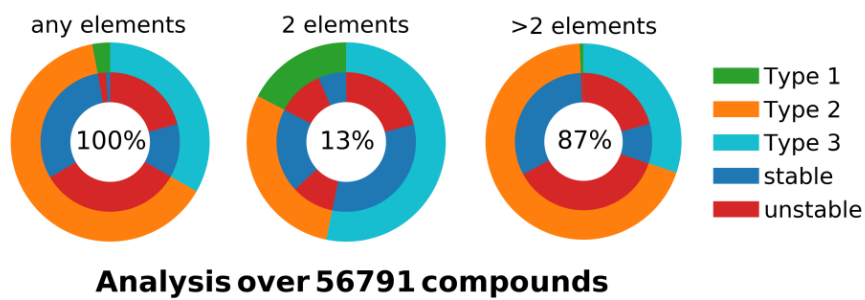

**Figure 2. Prevalence of reactions among known materials** Partitioning the compounds tabulated in Materials Project data into each of the three decomposition reaction types (outer circle). Then, for each type, partitioning compounds as stable (on the convex hull) and unstable (above the convex hull). Left – the entire database of 56,791 compounds; Center – only binary compounds; Right – only non-binary compounds. The fraction of the Materials Project comprising each circle is shown in the interior of each diagram.

**Functional performance on formation enthalpy predictions**

The decomposition reactions determining compound stability that are Type 1 are the least prevalent among Materials Project compounds (~3%) suggesting that $\Delta H_d$ rarely equals $\Delta H_f$, especially for $N > 2$ (< 1% of these compounds). Despite this, the primary approach currently used to benchmark first-principles thermodynamics methods is to compare experimental and computed $\Delta H_f$. We compared experimentally obtained $\Delta H_f$ from FactSage[23] to computed $\Delta H_f$ using the generalized gradient approximation (GGA) density functional as formulated by Perdew, Burke, and Ernzerhof (PBE)[24] and using the strongly constrained and appropriately normed (SCAN)[25] meta-GGA density functionals for 1,012 compounds spanning 62 elements (see **Fig. S1** for the prevalence of each element in the evaluated compounds). Importantly, this reduced space of compounds with experimental thermodynamic data decompose into the full range of Type 1 (37%), 2 (22%), and 3 (41%) reactions. However, we first only analyzed $\Delta H_f$ for all compounds to establish a baseline for subsequent comparison to $\Delta H_d$. On this set of 1,012 compounds, the mean absolute difference (MAD) between experimentally determined $\Delta H_f$ (at 298 K)[23] and calculated $\Delta H_f$, nominally at 0 K and without zero-point energy (ZPE), was found to be 196 meV/atom for PBE and 88 meV/atom for SCAN (**Fig. 3**a). In addition to a reduction in the magnitude of residuals by ~55%, the distribution of residuals is nearly centered about 0 for SCAN in contrast to PBE which consistently understabilizes compounds relative to their constituent elements (particularly diatomic gases), leading to predictions of $\Delta H_f$ that are too positive by ~200 meV/atom. Unlike PBE, SCAN has been shown to perform well for a range of diversely bonded systems[25-27] and does not suffer from this same systematic error. To probe this elemental dependence, the MAD for $\Delta H_f$ is partitioned for various chemical subsets of the dataset in **Fig. S2**. The performance of PBE is considerably worse for compounds containing gaseous elemental phases (MAD = 250 meV/atom) than for all other compounds (MAD = 138 meV/atom). This is in contrast to SCAN which performs slightly better when gaseous elements are present (MAD = 78 meV/atom) than for all other compounds (MAD = 99 meV/atom). The larger MAD associated with the latter set may be attributed to the increased prevalence of transition metals when gaseous elements are not present. We find the MAD for SCAN increases from 71 meV/atom for 489 compounds without transition metals to 103 meV/atom for 523 compounds with one or more transition metal. PBE does not exhibit this chemical dependence with large MAD of 197 meV/atom and 195 meV/atom for compounds with and without transition metals.

The near zero-centered residuals produced by SCAN suggest that no global systematic difference likely exists between the energies predicted by this density functional and those obtained experimentally, and thus, some of the lingering disagreement may arise from deficiencies in the functional for describing certain types of compounds, e.g. those with transition metals,[27-30] and/or be related to correlated noise in experimental measurement. For 228 binary and ternary compounds reported in Ref. 3 (compiled from Ref. [31]), the MAD between experimental sources (i.e., Refs. 23 and [31]) for $\Delta H_f$ is 30 meV/atom (**Fig. S3**). This difference agrees well with the scale of chemical accuracy expected for the experimental determination of $\Delta H_f$ of ~1 kcal/mol (~22 meV/atom for binary compounds)[27] and suggests that the disagreement between experiment and theory should not be lower than ~30 meV/atom on average because this is the magnitude of uncertainty in the experimental determination of $\Delta H_f$.

A potential source of disagreement between experimentally obtained and DFT-calculated $\Delta H_f$ is the incongruence in temperature, where experimental measurements are performed at 298 K and DFT calculations of $\Delta H_f$ are computed at 0 K, typically neglecting the effects of heat capacity from 0 K to 298 K as well as ZPE. These contributions are typically assumed to be small based on the results obtained for a limited set of compounds.[32] This assumption is robustly confirmed here for 647 structures where the vibrational and heat capacity effects on $\Delta H_f$ are shown to be ~7 meV/atom on average at 298 K (**Fig. S4**). Notably, at higher temperatures, the effects of entropy are significant and should be considered for accurate stability predictions at elevated temperature.[33]

**Optimizing elemental reference energies**

Various approaches have been developed to improve the PBE prediction of $\Delta H_f$ by systematically adjusting the elemental energies, $E_i$, of some or all elemental phases.[2,3,8-10] In the fitted elemental reference energy scheme, the difference between experimentally measured and calculated $\Delta H_f$ is minimized by optimally adjusting $E_i$ by a correction term, $\delta\mu_i$:

$$\Delta H_{f, A_{\alpha_1} B_{\alpha_2} \ldots} = E_{A_{\alpha_1} B_{\alpha_2} \ldots} - \sum_i \alpha_i (E_i + \delta\mu_i). \qquad [3]$$

To quantify the magnitude of errors that can be resolved by adjustments to the elemental reference energies, we applied **Equation 3** to $\Delta H_f$ computed with PBE and SCAN (**Fig. 3**b) with all elements considered in this optimization (these approaches are denoted in this work as PBE+ and SCAN+, respectively). Fitting reference energies for PBE approximately halves the difference between experiment and calculation and centers the residuals (MAD = 100 meV/atom). Because the difference between experiment and SCAN is less systematic, fitting reference energies improves SCAN errors substantially less than it improves PBE, and only reduces the MAD by ~20% (MAD = 68 meV/atom).

While adjusting elemental reference energies is simple and effective in reducing the difference between experimentally determined and calculated $\Delta H_f$ when density functionals produce systematic errors in the energies of the elemental phases, there are a number of limitations to this approach. Because it is a fitting scheme, the optimized $\delta\mu_i$ are sensitive to the set of experimental and calculated data used for fitting and do not necessarily have physical meaning, i.e., $\delta\mu_i$ accounts for the systematic disagreement between a density functional and experimental measurement across different types of materials, yet this can be difficult to interpret. Furthermore, the fitted reference energy scheme, as implemented here, produces a single $\delta\mu_i$ for each element whether a given element appears in the compounds as a cation or anion (e.g., $Sb^{3+}$ or $Sb^{3-}$). For the majority of the compounds considered in this work, the use a single fitted value is appropriate because elements only appear in the data as either anions or cations. However, if one was interested in studying compounds containing elements that appear as cationic or anionic, statistically resolving a separate $\delta\mu_i$ for cation- and anion-specific use would be more appropriate, as the fitted correction can differ in both magnitude and sign for cations and anions. Additionally, fitted reference energies have only been available for PBE (and for SCAN, as reported in this work), so the calculation of $\Delta H_f$ using alternative functionals which may be better suited for a given problem would require a re-fitting of reference energies within that functional. These limitations make it advantageous to avoid fitted reference energies for the high-throughput prediction of stability, particularly if they have negligible effect on the validity of first-principles predictions.

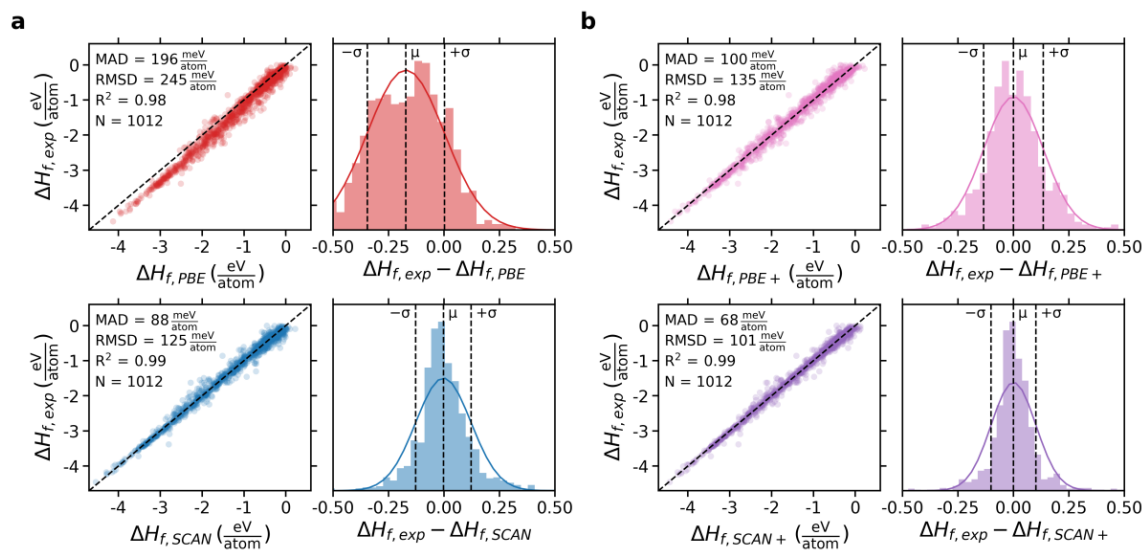

**Figure 3. Experimental vs. theoretical formation enthalpies (Type 1) a)** A comparison of experimentally measured and DFT-calculated $\Delta H_f$ for all 1,012 compounds analyzed (PBE above; SCAN below) showing that SCAN significantly improves the prediction of $\Delta H_f$ over PBE. MAD is the mean absolute difference; RMSD is the root-mean-square difference; $R^2$ is the correlation coefficient; N is the number of compounds shown; μ is the mean difference; σ is the standard deviation. A normal distribution constructed from μ and σ is shown as a solid curve. **b)** For the same compounds, a comparison of PBE and SCAN with experiment using fitted elemental reference energies for the calculation of $\Delta H_f$ (PBE+ above; SCAN+ below) showed that for Type 1 reactions fitted elemental reference energies significantly improve the prediction of $\Delta H_f$, especially predictions by PBE. These results are provided in **Table S1** (for elemental energies) and **Table S2** (for compound data). The chemical dependence of these results is shown in **Fig. S2**a and the distribution of $\Delta H_{f,exp}$ is provided in **Fig. S5**a.

**Decomposition reaction analysis**

While the improved construction of the SCAN meta-GGA density functional and the use of fitted reference energies ameliorates errors associated with the insufficient description of the elements and thus improves the prediction of $\Delta H_f$ considerably relative to PBE, the effects these approaches have on the prediction of thermodynamic stability – i.e., $\Delta H_d$ – have not yet been quantified. We used $\Delta H_f$ obtained from experiment, PBE, and SCAN for the 1,012 compounds analyzed in **Fig. 3** to perform the *N*-dimensional convex hull analysis to determine the decomposition reaction and quantify $\Delta H_d$. For 646 compounds that decompose by Type 2 or 3 reactions, the MAD between experimentally measured and DFT-computed $\Delta H_d$ is substantially lower than for $\Delta H_f$ – ~60% lower for PBE and ~30% lower for SCAN (**Fig. 4**). Notably, the decomposition reaction that results from using experiment, PBE, or SCAN is identical in terms of the competing compounds and their amounts for 89% of the 1,012 compounds evaluated.

For 231 Type 2 decomposition reactions where compounds compete only with compounds and fitted reference energies thus have no influence on $\Delta H_d$, SCAN and PBE are found to perform comparably with MADs of ~35 meV/atom compared with experiment. This agreement between theory and experiment using either functional approaches the "chemical accuracy" of experimental measurements (~1 kcal/mol = 22 meV/atom for binary compounds) and is similar to the difference in $\Delta H_f$ between two experimental sources evaluated in this work (30 meV/atom). A previous study of the formation energies of 135 ternary metal oxides from their constituent binary oxides found that PBE with a Hubbard $U$ correction specifically fit for transition metal oxides achieved a MAD of 24 meV/atom with experiment.[11] The formation of compounds with greater than two elements (ternaries, quaternaries, etc.) from their corresponding binaries is sometimes used as an approximation for $\Delta H_d$.[34,35] The energy of this reaction, $E_f^{binaries}$, is equivalent to $\Delta H_d$ when only elements and binary compounds are present in the decomposition reaction, but this becomes less likely as the number of competing compounds in a given chemical space increases. Our analysis of the Materials Project shows that compounds composed of > 2 elements are relevant in the decomposition reactions of 42% of 28,884 ternary compounds and 91% of 14,123 quaternary compounds. For these cases, $E_f^{binaries}$ does not equal $\Delta H_d$. As a demonstration of the magnitude of this disagreement, we selected four quaternary garnet oxides ($C_3A_2D_3O_{12}$) in our dataset (A = Al, D = Si, C = Ca/Mg/Mn/Fe) and found that $E_f^{binaries}$ overestimates stability (is more negative than $\Delta H_d$) by 69 meV/atom on average (see **Supporting Information** for more details). In **Fig. 4**, our results show excellent agreement between experiment and theory for $\Delta H_d$ of a diverse set of materials, considering all possible decomposition products and without requiring a Hubbard $U$ correction. Because Type 2 decomposition reactions only involve compounds, computing the decomposition reaction energy using total energies or formation enthalpies is equivalent – therefore the results with (**Fig. 4**b) and without (**Fig. 4**a) fitted reference energies are identical.

Elemental energies are included in the calculation of $\Delta H_d$ for compounds that compete thermodynamically with both compounds and elements (Type 3 decomposition reactions). However, for 415 reactions of this type and using either SCAN or PBE we found that the use of fitted reference energies does not significantly affect the agreement with experiment for $\Delta H_d$ with improvements of only ~2 meV/atom (**Figs. 4**c, d). For these compounds, SCAN improves upon PBE by ~20% and the MAD between SCAN and experiment (73 meV/atom) falls between those for Type 1 (88 meV/atom) and Type 2 (34 meV/atom) reactions.

The prevalence of each reaction type was quantified for the Materials Project database, with Type 2 reactions accounting for 63% of all decompositions evaluated and this fraction increasing from 29% to 67% to 75% for binary, ternary, and quaternary compounds, respectively. For these cases, our results show that both SCAN and PBE can be expected to yield chemically accurate predictions of $\Delta H_d$, which quantifies the driving force for thermodynamic stability. While on average, SCAN and PBE perform similarly for $\Delta H_d$, this analysis is performed only on ground-state structures within each functional. It was recently shown that SCAN performs significantly better than PBE for structure selection – i.e., identifying the correct polymorph ordering of which crystal structure is the lowest energy at fixed composition.[27] Here, ~10% of the 2,238 structures optimized were found to have different space groups using PBE and SCAN. Considering only ground-states, the lowest energy PBE and SCAN structures differ for ~11% of the 1,012 unique compositions assessed in this work. While the MAD from experiment for $\Delta H_d$ calculated by SCAN and PBE differs by only ~20%, additional advantages are likely associated with the use of SCAN for the accurate description of structure and properties.[25-27,36] The discrepancies between the structures and polymorph energy orderings predicted by PBE and SCAN with experiment may also contribute to the reported differences between the approaches.

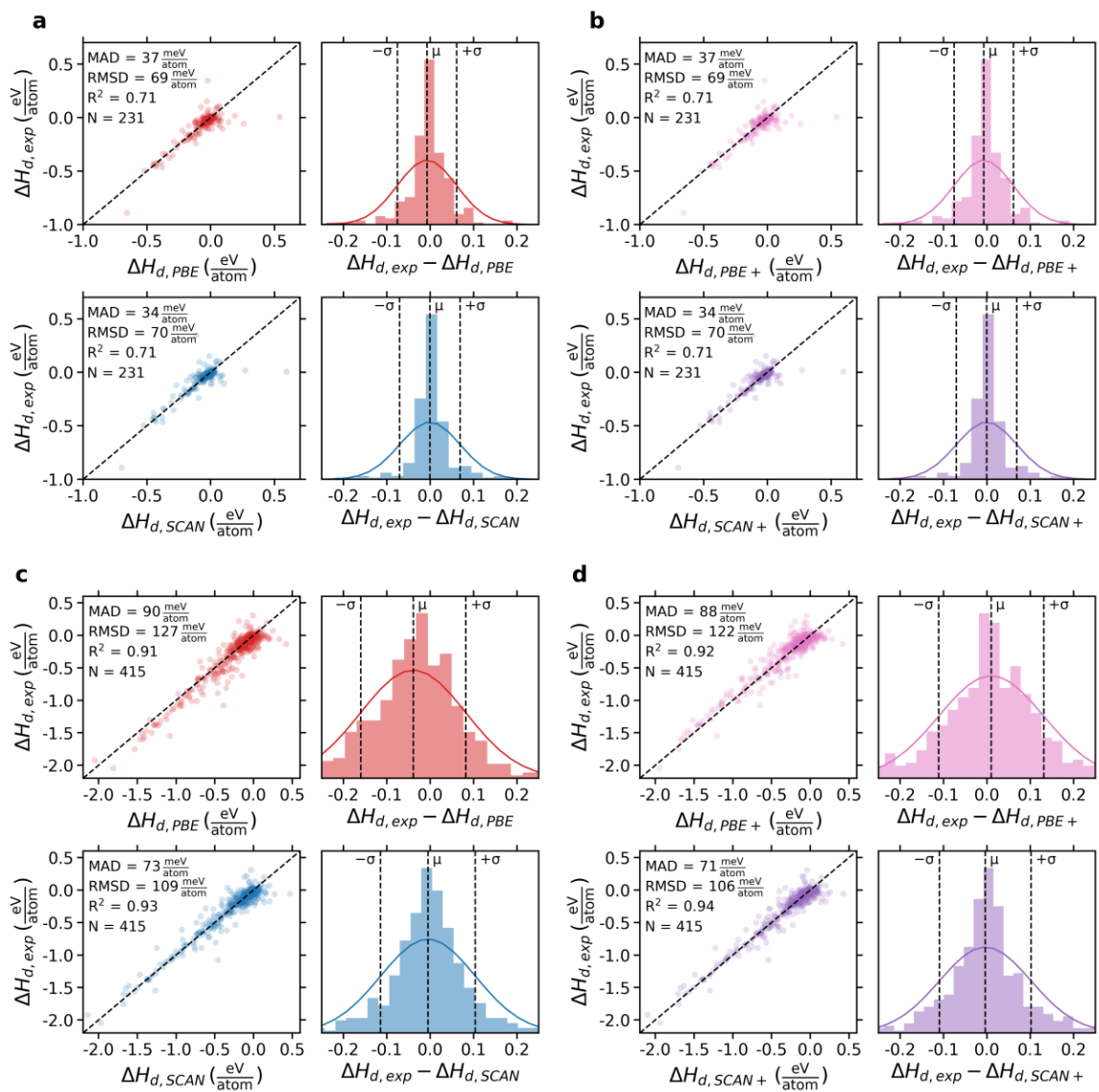

**Figure 4. Experimental vs. theoretical decomposition enthalpies a)** A comparison of experimentally measured and DFT-calculated $\Delta H_d$ (PBE above; SCAN below) for 231 compounds that undergo Type 2 decomposition reactions showing similar performance between PBE and SCAN in predicting $\Delta H_d$. The chemical dependence of the MAD between theory and experiment for Type 2 reactions is shown in **Fig. S2**b and the distribution of $\Delta H_{d,exp}$ for Type 2 reactions is provided in **Fig. S5**b. **b)** For the same compounds, a comparison of PBE and SCAN with experiment using fitted elemental reference energies for the calculation of $\Delta H_d$ (PBE+ above; SCAN+ below) showing identical results as **(a)** due to a cancellation of elemental energies for these Type 2 decomposition reactions. **c)** A comparison of experimentally measured and DFT-calculated $\Delta H_d$ (PBE above; SCAN below) for 415 compounds that undergo Type 3 decomposition showing similar performance between PBE and SCAN in predicting $\Delta H_d$. **d)** For the same compounds, a comparison of PBE and SCAN with experiment using fitted

elemental reference energies for the calculation of $\Delta H_d$ (PBE+ above; SCAN+ below) showing that including fitted elemental reference energies does not significantly improve the prediction of $\Delta H_d$ for Type 3 decomposition reactions. Annotations are as described in the **Fig. 3** caption. The chemical dependence of the MAD between theory and experiment for Type 3 reactions is shown in **Fig. S2**c and the distribution of $\Delta H_{d,exp}$ for Type 3 reactions is provided in **Fig. S5**c.

## DISCUSSION

For 1,012 compounds, we show that fitting elemental reference energies for both GGA (PBE) and meta-GGA (SCAN) density functionals improves computed formation enthalpies, $\Delta H_f$ (**Fig. 3**). However, to accurately predict the stability of materials, it is essential to accurately compute the decomposition enthalpy, $\Delta H_d$, which dictates stability with respect to all compounds and elements in a given chemical space. $\Delta H_d$ is computed by determining the stoichiometric decomposition reaction with the most positive reaction energy. $\Delta H_f$ is only relevant for the stability of compounds that undergo Type 1 decompositions, where the compound only competes with elemental phases and consequently, $\Delta H_d = \Delta H_f$. (**Fig. 1**). Furthermore, Type 1 decompositions occur for only 17% of binaries and almost never (< 1%) for non-binaries, as shown for the ~60,000 *N*-component compounds evaluated (**Fig. 2**). For this reason, $\Delta H_f$ and the agreement between experiment and theory for $\Delta H_f$ are rarely relevant to the stability of materials. However, for other applications such as the calculation of defect formation energies, $\Delta H_f$ is the relevant materials property and the adjustment of calculated chemical potentials using the fitted elemental reference energy scheme may still have significant utility, especially when using PBE. The accuracy of $\Delta H_f$ is also critical when only select compounds in a given chemical space are not well-described by a given functional – e.g., when calculating the stability of peroxides with PBE and the correction developed by Wang *et. al*,[9] where $O_2^{2-}$ groups are overstabilized.[11,37] If a given error is not systematic for all compounds in a given chemical space, errors in $\Delta H_f$ may propagate to the errors in $\Delta H_d$.

The stabilities of compounds that undergo Type 2 decompositions (63% of compounds tabulated in Materials Project) can be determined without any consideration of elemental energies. For these compounds, PBE and SCAN perform similarly and approach the resolution of experimental approaches to determining $\Delta H_f$ (~30 meV/atom) (**Fig. 4**a, **Fig. S3**). Importantly, the performance metrics we provide are evaluated over a wide range of compounds and chemistries. For chemical spaces that are known to be problematic for a given approach (e.g., 3*d* transition metals for PBE), the error can significantly exceed the average difference reported here.[27,30]

While the majority of compounds in the Materials Project compete with Type 2 decomposition reactions, this is not generally known when first evaluating a compound and so high-throughput screening approaches that typically survey a wide range of compounds will likely include analysis of Type 1 and Type 3 decomposition reactions that do require the calculation of elemental energies. Type 1 decompositions, which occur for binary compounds in sparsely explored chemical spaces, will be highly sensitive to the functional and elemental energies and SCAN improves significantly upon PBE for these compounds. Notably, fitting elemental reference energies for PBE still results in larger errors than SCAN and fitting reference energies for SCAN leads to only modest additional improvements. For Type 3 decompositions, which are ~10× more prevalent than Type 1 reactions in Materials Project, SCAN improves upon PBE by ~20% and the use of fitted elemental reference energies has almost no effect (~2 meV/atom on average) on either approach (**Figs. 4**c-d). Interestingly, considering the ~60,000 compounds in Materials Project (**Fig. 2**, left), a roughly equal fraction of Type 2 compounds are stable (48%) and unstable, yet only 37% of Type 3 compounds are stable. However, Type 3 compounds are more amenable to non-equilibrium synthesis approaches that allow for increased chemical potentials of the elements and thus potential access to metastable compounds.[22]

In summary, we've shown that the decomposition reactions that dictate the stability of solid compounds can be divided into three types that depend on the presence of elemental phases in the decomposition reaction. Through a global evaluation of phase diagrams for ~60,000 compounds in the Materials Project, we quantify the prevalence of these reaction types and show that the formation enthalpy is rarely the quantity of interest for stability predictions (~3% of Materials Project compounds). Instead, the decomposition enthalpy, which may or may not include the calculation of elemental phases is the most relevant quantity. Benchmarking the PBE and SCAN density functionals against decomposition enthalpies obtained from experimental data reveals quantitatively and qualitatively different results than benchmarking only against formation enthalpies and in most cases mitigates the need to systematically correct DFT-calculated elemental energies for the assessment of stability.

We showed that for 231 reaction energies between compounds, the agreement between SCAN, PBE, and experiment (~35 meV/atom) is comparable to the expected noise in experimental measurements. The differences between experiment and theory are systematically lower for $\Delta H_d$ than for $\Delta H_f$ no matter the choice of functional or elemental reference energies. This can be attributed to cancellation of errors within a given chemical space (phase diagram). For example, if we consider the stability of fluorides calculated with PBE, $\Delta H_f$ will be too positive for all fluorides competing for stability with one another because PBE over-stabilizes the $F_2$ reference state. However, because this systematic overestimation of $\Delta H_f$ often persists for all compounds in the decomposition reaction, the energy of that decomposition reaction, $\Delta H_d$ usually agrees considerably better with experiment than $\Delta H_f$. SCAN does not suffer from this same systematic error with respect to diatomic gaseous elemental reference states, though it is plausible that some lingering error persists in the SCAN description of dissimilar systems (e.g., metals and insulators) as is often present in the calculation of $\Delta H_f$. Nevertheless, the compounds that compete for stability are typically much more chemically similar to one another than they are to their constituent elemental reference states, leading to a more consistent description of the energies required to calculate $\Delta H_d$ than $\Delta H_f$. In **Fig. S2**, the agreement between each functional and experiment is shown for various chemical subsets of the data (oxides, halides, etc.). In this analysis, we find that while the prediction of $\Delta H_f$ is highly sensitive to the chemical composition for PBE and moderately sensitive for PBE+, SCAN, and SCAN+, the prediction of $\Delta H_d$ for Type 2 reactions varies minimally for each functional as the chemical composition is varied. Therefore, because this type of decomposition reaction is predominant in determining solid stability, we show that high-throughput DFT approaches to stability predictions are generally in excellent agreement with experiment for a diverse set of materials. For alternative decomposition reactions that include both compounds and elements or problems that require higher energy resolution such as polymorph energy ordering,[28,36] the choice of functional (e.g., SCAN instead of PBE) can have non-negligible effects on stability predictions.

## METHODS

Experimental values for $\Delta H_f$ were obtained from the FactSage database[23] for 1,012 compounds as reported at 298 K and 1 atm. For each compound, the NREL Materials Database (NRELMatDB)[3] was queried for structures matching the composition within 50 meV/atom of the ground-state structure as reported in the database. If a given compound had no calculated structures tabulated in NRELMatDB, the procedure was repeated with the Materials Project database[1]. Structures containing potentially magnetic elements were sampled in non-magnetic, two ferromagnetic (high- and low-spin), and up to 16 antiferromagnetic configurations (depending on cell configuration) where the ground-state magnetic configuration was retained for each structure. Sampling was performed using the approach described by NRELMatDB. This process was also repeated for all 62 elements represented in the dataset with the exceptions of $H_2$, $N_2$, $O_2$, $F_2$, and $Cl_2$ which were calculated as diatomic molecules in a $15\times15\times15$ Å box. After magnetic sampling, 2,238 unique structures were found for the 1,012 compounds and 62 elements. All structures were optimized with PBE and SCAN using the Vienna Ab Initio Simulation Package (VASP)[38,39] using the projector augmented wave (PAW) method[40,41], a plane wave energy cutoff of 520 eV, and a Γ-centered Monkhorst-Pack k-point grid with $20|b_i|$ discretizations along each reciprocal lattice vector, $b_i$. The energy cutoff, k-point density, and related convergence settings were sufficient to achieve total energy convergence of < 5 meV/atom for all calculations. Pseudopotentials used for each element are provided in **Table S1**. For the calculation of phonons to compute thermal effects, the finite displacement method with $2\times2\times2$ supercells as implemented in PHONOPY[42] was used with SCAN and an increased plane wave cutoff of 600 eV and further tightened convergence criteria for total energy convergence of < 1 meV/atom. These results are compiled in **Table S3**.

**Data availability**

All necessary data for reproducing this work is contained within, including the computed formation enthalpies, decomposition enthalpies, and chemical potentials that are provided in the Supplementary Information. Additional data are available from the corresponding author upon request.

**Acknowledgements**
The authors gratefully acknowledge Vladan Stevanović for many useful discussions on the thermodynamics of solids and thank Christopher Sutton and Jacob Clary for providing feedback on the manuscript. This work was supported by the National Science Foundation (award nos. CBET-1433521, CHE-1800592, and CBET-1806079). The authors also acknowledge partial support for this work from the U.S. Department of Energy, Office of Basic Energy Sciences (S.L. and A.M.H., contract no. DE-AC36-08GO28308) and Fuel Cell Technologies Office (A.W.W. and C.B.M, award no. DE-EE0008088). High Performance Computing resources were sponsored by the U.S. Department of Energy's Office of Energy Efficiency and Renewable Energy, located at NREL.

**Contributions**
C.J.B. performed the calculations and analysis and drafted the manuscript. A.W.W provided the experimental dataset and assisted in its curation. S.L., C.B.M. and A.M.H supervised the project, provided guidance on the calculations and analysis, and assisted in writing the manuscript. All authors commented on the results and reviewed the manuscript.

**Competing interests**
The authors declare no competing interests.

**Corresponding authors**
Correspondence to Charles Musgrave (charles.musgrave@colorado.edu) or Aaron Holder (aaron.holder@colorado.edu)